\documentclass{article}
\usepackage{icmc2026_paper_template}
\usepackage{times}
\usepackage{ifpdf}
\usepackage{subfig}
\usepackage{floatrow}
\usepackage{soul}
\usepackage{url}
\usepackage[english]{babel}

\def\papertitle{Music of Changing Lines: Toward a Culturally Situated Approach to the I-Ching}
\def\firstauthor{Ling Qi}
\def\secondauthor{Aleksandra Teng Ma}
\def\thirdauthor{Alexandria Smith}

\newif\ifpdf
\ifx\pdfoutput\relax
\else
   \ifcase\pdfoutput
      \pdffalse
   \else
      \pdftrue
  \fi
\fi

\ifpdf 
  \usepackage[pdftex,
    pdftitle={\papertitle},
    pdfauthor={\firstauthor, \secondauthor, \thirdauthor},
    bookmarksnumbered, 
    pdfstartview=XYZ 
   ]{hyperref}

  \usepackage[pdftex]{graphicx}
  \graphicspath{{./figures/}}
  \DeclareGraphicsExtensions{.pdf,.jpeg,.png}

  \usepackage[figure,table]{hypcap}

\else 
  \usepackage[dvips,
    bookmarksnumbered, 
    pdfstartview=XYZ 
  ]{hyperref}  

  \usepackage[dvips]{epsfig,graphicx}
  \graphicspath{{./figures/}}
  \DeclareGraphicsExtensions{.eps}

  \usepackage[figure,table]{hypcap}
\fi

\hypersetup{
    colorlinks,%
    citecolor=black,%
    filecolor=black,%
    linkcolor=black,%
    urlcolor=black
}

\title{\papertitle}

 \threeauthors
   {0.5in}
   {\firstauthor\textsuperscript{*}} {Georgia Institute of Technology \\ %
     {\tt \href{lqi60@gatech.edu}{lqi60@gatech.edu}}}
   {\secondauthor\textsuperscript{*}} {Georgia Institute of Technology \\ %
     {\tt \href{tengofma@gmail.com}{tengofma@gmail.com}}}
   {\thirdauthor} { Georgia Institute of Technology \\ %
     {\tt \href{alexandria.smith@gatech.edu}{alexandria.smith@gatech.edu}}}

\makeatletter
\g@addto@macro\@author{%
  \\[4ex]
  {\noindent
  \centering
  \footnotesize
  \textsuperscript{*}These authors contributed equally to this work.\par
  }
}
\makeatother

\begin{document}
\capstartfalse
\maketitle
\capstarttrue
\begin{abstract}
The \textit{I-Ching} is one of the most influential texts in Chinese intellectual history, integrating divination, cosmology, and ethical reflection. While Western experimental music, most notably John Cage, has drawn on the \textit{I-Ching} as a source of chance operation, such appropriations have often detached its formal mechanisms from the interpretive and philosophical processes that give the text meaning. This work, \textit{Music of Changing Lines}, presents an interactive system that re-centers the \textit{I-Ching} as a meaning-bearing framework rather than a neutral randomizer. Users perform Wen Wang Fa coin casting, which is accompanied in real time through probabilistic musical processes. The resulting hexagrams and changing lines are interpreted by a large language model, Gemini, in relation to the user’s inquiry. This textual interpretation is then translated into a prompt for a generative music model, Lyria, producing a responsive musical realization. By situating AI as an interpretive intermediary rather than a compositional authority, the system foregrounds the \textit{I-Ching}’s ritual, interpretation, and participation as the primary sonic materials. \textit{Music of Changing Lines} extends process-driven traditions in computer music by demonstrating how generative AI can support participatory, meaning-driven musical processes without prescribing musical structure or replacing human agency.
\end{abstract}

\section{Introduction}\label{sec:introduction}
\begin{figure}[t]
  \centering
  \includegraphics[width=\columnwidth]{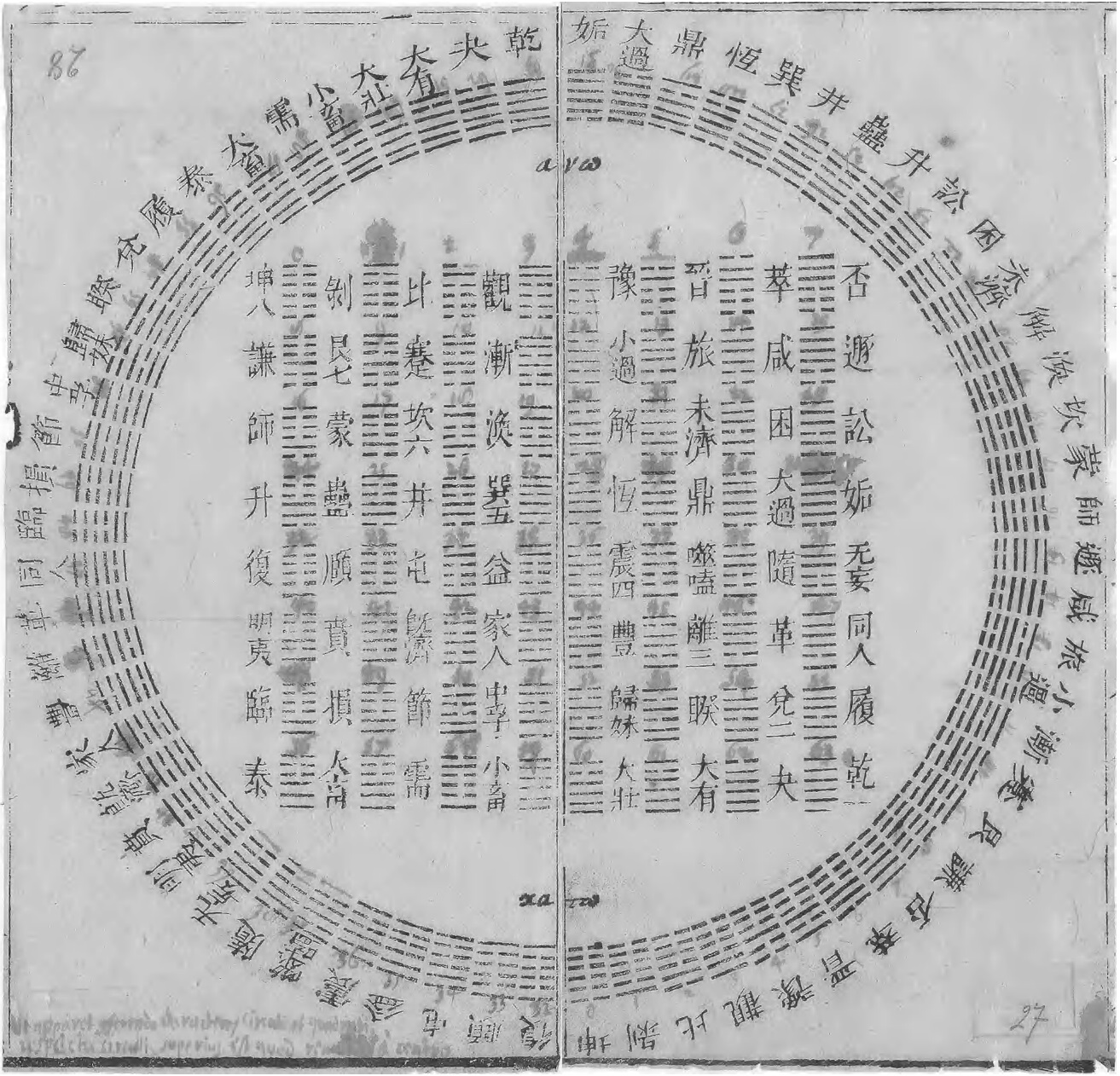}
  \caption{Hexagrams of the \textit{I-Ching} that belongs to German mathematician philosopher Gottfried Wilhelm Leibniz \cite{Perkins_2004}.}
  \label{fig:iching}
\end{figure}

The \textit{I-Ching} \cite{yijingWilhelm}, or \textit{Book of Changes}, has circulated for more than three millennia as both a divinatory system and a philosophical text. Its hexagrams (Figure~\ref{fig:iching}), generated through ritualized chance procedures such as the Wen Wang Fa coin-toss method, provide a framework for interpreting uncertainty, change, and the unfolding of events. Through centuries of commentary and reinterpretations, most notably within the Confucian tradition, the text’s symbols have been read as both oracular signs and ethical or metaphysical principles. 

In the twentieth century, the \textit{I-Ching} entered Western experimental music most prominently through the work of John Cage. Cage was introduced to the \textit{I-Ching} by Christian Wolff, whose father published its English translation \cite{kostelanetz2003conversing}. 
Drawn to the use of chance, Cage recontextualized the \textit{I-Ching} as a compositional tool in works such as \textit{Music of Changes} \cite{CageMusicChanges} and \textit{Music for Piano 21–52} \cite{cage1960_music_for_piano}. 
He used coin tosses to form hexagrams, which then determined values for musical parameters such as sound type, duration, and dynamics. 
Despite Cage's prior personal engagement with the text as a source of philosophical and life guidance \cite{cage1961silence}, this approach largely set aside the interpretive and cosmological readings of the hexagrams that traditionally follow the casting process.
Hence, while these works marked the first explicit introduction of the \textit{I-Ching} into Western experimental music, they deviated from the \textit{I-Ching}'s philosophical and divinatory context. 

\begin{figure*}[t]
  \centering
  \includegraphics[width=\textwidth]{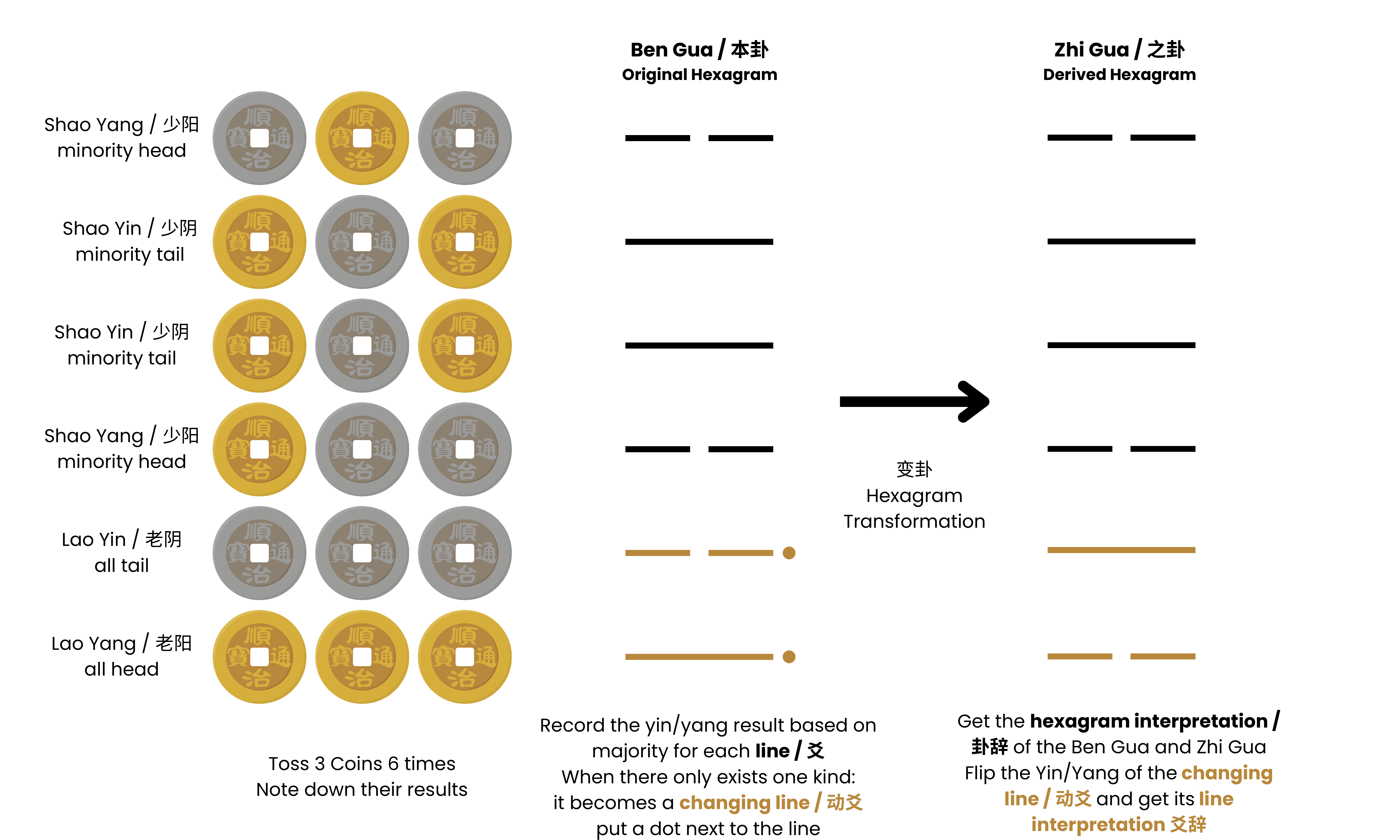}
  \caption{Overview of the Wen Wang Fa coin tossing approach in the \textit{I-Ching}.}
  \label{fig:casting}
\end{figure*}

In this work, \textit{Music of Changing Lines}, we revisit the \textit{I-Ching} not merely as a mechanism for randomness but as a culturally situated system of meaning. Our creative goal is to design an immersive, interactive divination experience that translates the full casting and interpretive process of the \textit{I-Ching} into a participatory sonic ritual. Rather than generating music solely from rule-based mappings, we integrate the divinatory semantics of the \textit{I-Ching}: the Ben Gua (original hexagram), Zhi Gua (derived hexagram), and any Dong Yao (changing lines) guide the musical affect, structure, and evolution.
To realize the personalized interpretation required by the \textit{I-Ching} process, we draw on contemporary machine learning tools for text and music generation. A large language model (Gemini 2.5 Flash \cite{team2023gemini, comanici2025gemini}) interprets the hexagrams in relation to the user’s question, synthesizing elements from the Gua Ci (hexagram interpretation) and Yao Ci (line interpretation). This interpretation is then translated into musical descriptors, such as mood, texture, pacing, instrumentation, which condition a real-time generative music model (Lyria \cite{team2025live}) to produce a responsive sonic representation of the divination outcome. The casting stage retains Cagean chance operations, while the interpretive stage reintroduces the symbolic and philosophical dimensions that Cage bracketed out.

Through this work, we explore how ancient divination practices can be reimagined as interactive music systems, and how computer music AI may function as an interpretive intermediary within a participatory process, instead of a generator of finished artifacts.
Rather than approaching the \textit{I-Ching} as a compositional mechanism in the Cagean sense, this work treats computer music as a medium to accompany the divination process. 
Here, process---rather than musical output---becomes the primary compositional object, and sound functions as a means of articulating uncertainty, interpretation, and becoming.

\section{Related Works}\label{sec:relatedWorks}
\subsection{John Cage and Chance Operation}\label{subsec:cage}
John Cage’s \textit{Music of Changes} \cite{CageMusicChanges} stands as a foundational work in the history of chance-based composition. 
In this work, Cage constructed charts for musical parameters like tempo, duration, dynamics, density, and sound type. 
Each chart consists of 64 values, corresponding to the \textit{I-Ching}'s 64 hexagrams. 
Cage used repeated coin tosses to form hexagrams and let chance select values from the charts at each structural point of the composition. 
His compositional goal was to eliminate personal taste and intention, allowing sound to emerge “free of individual memory and psychology” \cite{cage1961silence}. 
Before Cage, postwar avant-garde music was largely defined by serialist and modernist systems that sought total compositional control.
Cage's chance operation marked an important turn within twentieth-century experimentalism, influencing Fluxus, Neo-Dada, and later developments in algorithmic and stochastic music \cite{brecht1963anthology, sitsky2002music, joseph2016experimentations}.

Scholars have since debated the implications of Cage’s technique \cite{patterson1996appraising, jensen2009john, bernstein2010writings}.
Jensen \cite{jensen2009john} argues that despite Cage's genuine fascination with Eastern philosophy, he employed a Westernized, modernist reading of the \textit{I-Ching} that emphasized randomness over its traditional semantic and divinatory meanings.
Furthermore, Jensen suggests that Cage’s procedures parallel the ``chaos game,'' generating forms that are unpredictable yet still governed by underlying constraints, illustrating that Cage’s rejection of intention does not necessarily yield pure indeterminacy.

\subsection{The \textit{I-Ching} and Its Westernization}\label{subsec:iching}
To address the cultural and epistemological gap between Cage’s procedural adoption and the \textit{I-Ching}’s original function, it is important to situate the text within its historical foundations. 
The \textit{I-Ching}, originally written during the Zhou dynasty (1000--750 BCE), serves both as a manual of divination and a philosophical text rooted in Daoist and Confucian cosmology.

Traditional \textit{I-Ching} divination systems such as Wen Wang Fa rely on coin tossing, a process we illustrated in Figure~\ref{fig:casting}. It begins by casting three coins six times, in order to generate a hexagram composed of yin (broken) and yang (solid) lines, along with any changing lines (\textit{Dong Yao}) that indicate transformation.
The original hexagram is called \textit{Ben Gua}, and the transformed hexagram derived from Dong Yao is called \textit{Zhi Gua}.
There are 64 possible hexagrams in total, each with a corresponding interpretation (\textit{Gua Ci}).
Ben Gua's Gua Ci provides an overarching judgment of the situation, while the \textit{Yao Ci} of any changing lines offers more fine-grained guidance about movement, transition, and the unfolding of events.
Interpreters then compare the \textit{Ben Gua} with the \textit{Zhi Gua} (the transformed hexagram derived from the changing lines) to discern their position within a shifting cosmological pattern. 

The translation of the \textit{I-Ching} (Princeton University Press, 3rd ed.) by German sinologist Richard Wilhelm, rendered into English by American Jungian psychologist Cary F. Baynes, is the most influential Western version of the ancient Chinese classic \cite{yijingWilhelm}. 
However, this edition reframes the text through a Western intellectual lens, emphasizing the \textit{I-Ching} as a source of personal introspection rather than a ritual system \cite{smith2012book}. 
This is the version that John Cage read, and it likely influenced his understanding of the \textit{I-Ching} as a method of pure chance, stripped of its original symbolism and semantic interpretations. 

\subsection{Other Musical Transcodings of the \textit{I-Ching}}\label{subsec:musical-iching}
Apart from John Cage, numerous composers have devised techniques that translate the structural or symbolic features of the \textit{I-Ching} into musical parameters \cite{chung1995ching, muller2016real, xue2019transcoding}.
Chou Wen-Chung’s “variable modes” partition the twelve-tone scale into three conjuncts---Heaven, Humans, and Earth---deriving melodic intervals from the cosmic significance of the hexagrams \cite{xue2019transcoding}.
Zhao Xiaosheng's Tai Chi framework draws on the yin–yang dynamics embedded in the hexagrams to generate pitch materials, producing cyclical musical motion analogous to hexagram transformation processes \cite{xue2019transcoding}.
Chung Yiu-Kwong's I-Ching Compositional System \cite{chung1995ching} connects the commentaries (Gua Ci) and individual line statements to pitch organization and the temporal statements trajectory of musical sections.
Jeremy Muller's research \cite{muller2016real} explores real-time chance procedures in digital contexts, demonstrating how contemporary technologies can replicate Cagean indeterminacy.
Collectively, these approaches reveal the wide range of musical possibilities inspired by the \textit{I-Ching}, ranging from symbolic mappings to technologically driven, interactive systems.

\begin{figure*}[t]
  \centering
  \includegraphics[width=\textwidth]{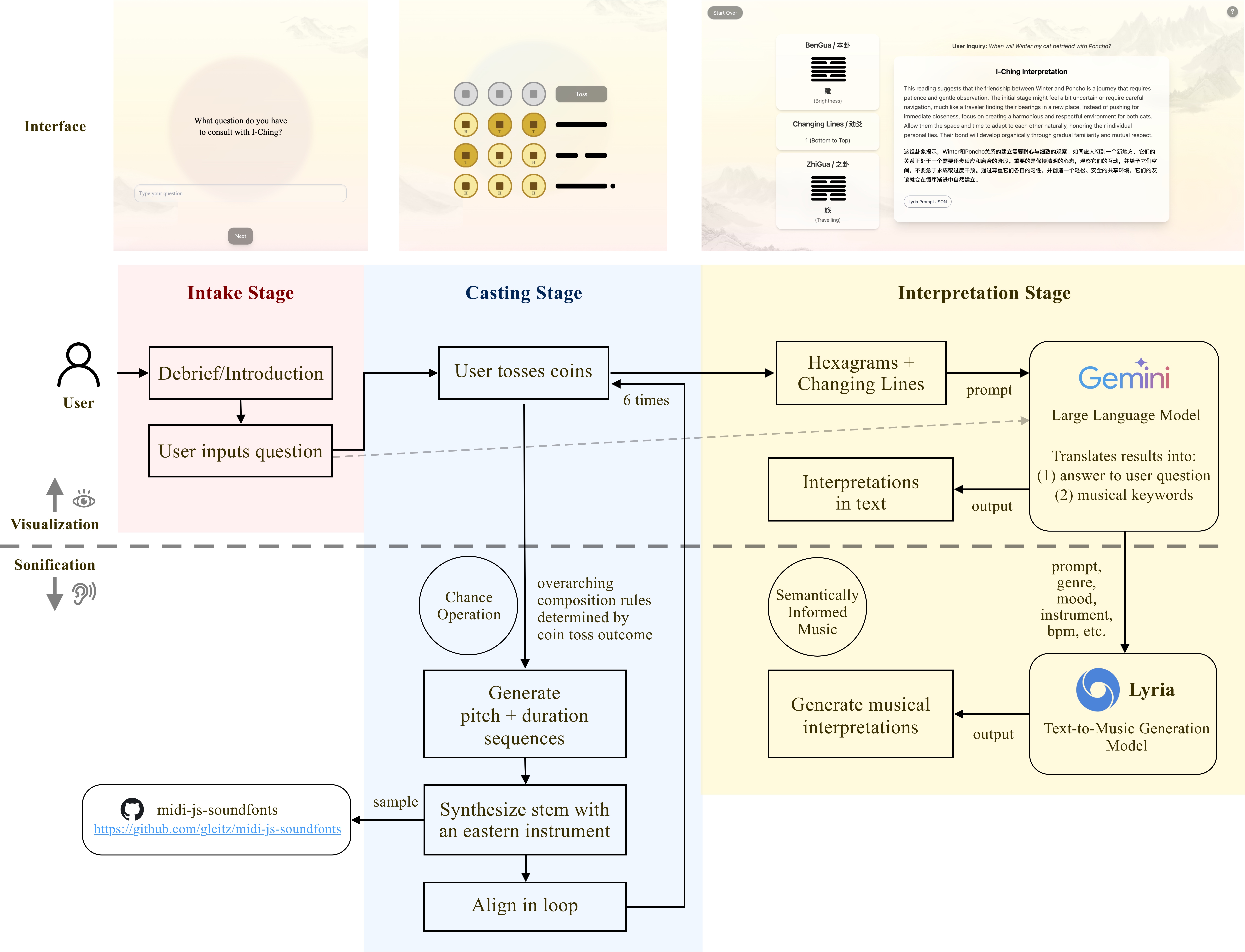}
  \caption{User Interface and Flow Chart for the system, \textit{Music of Changing Lines}}
  \label{fig:flowchart}
\end{figure*}

\section{System Design}
\textit{Music of Changing Lines}
\footnote{The source code of our application is available at 
\url{https://github.com/LingQi000809/iching-sonification}. A video walk-through of our system can be viewed at \url{https://vimeo.com/1150693113?share=copy&fl=sv&fe=ci}.} 
is implemented as a web application that aims to create an immersive, audio-visual experience of the \textit{I-Ching} divination. 
Users are guided through three primary stages: intake, casting, and interpretation.
The user interface and flow chart of our work are shown in Figure~\ref{fig:flowchart}.

\subsection{Intake Stage}\label{subsec:intake}
After being introduced to the system and the relevant context of the \textit{I-Ching} process, users are invited to submit a question for consultation, which may range from personal dilemmas to everyday choices.
While classical \textit{I-Ching} readings often draw on detailed personal information, our system only requests an optional name, preserving privacy while maintaining a sense of personalization.

\subsection{Casting Stage}\label{subsec:casting}
The Casting Stage operationalizes the chance component of the \textit{I-Ching} divination. 
Users perform six coin tosses, each consisting of three coins, to form a hexagram. 
To enhance engagement and personalize the experience, we animate the coin tosses and require the user to initiate each toss via an interactive button. 
The result is unique to each user and the moment of interaction, embodying the traditional notion that chance is intertwined with both the individual and the universe.

During casting, semantic interpretations are deliberately withheld, following the \textit{I-Ching} principles that forbid premature reading of individual lines. 
Because this stage carries no interpretive meaning, we use a rule-based, probability driven method, similar to John Cage's chance operation.
We sampled taiko drum, koto, shakuhachi, shamisen, flute, and nylon-string guitar (as an approximation of the pipa) from a soundfont library\footnote{https://github.com/gleitz/midi-js-soundfonts}, with each coin row mapped to one of these instruments.
The instruments were chosen from the library to provide the closest available match to traditional Chinese sounds.
After a toss, pitch and duration sequences are generated probabilistically for the corresponding instrument, then looped and layered with previous instruments to produce a cohesive, multi-voiced soundscape.
The probability is informed by the coin outcome: for example, a three-head toss (old yang) favors ascending, longer notes, whereas three tails (old yin) favors descending, shorter notes. 
This allows the music to reflect the dynamics of the universe at the moment of casting, without conveying semantic meaning. 
All melodies are confined to a pentatonic scale to maintain harmonic coherence and evoke an aesthetic commonly associated with East Asian musical traditions, consistent with the \textit{I-Ching} context.
Unlike Cage, who relied on pre-defined charts, we use probability distributions to give more agency to chance while keeping aesthetic constraints that make the music immersive rather than distracting.

\subsection{Interpretation Stage}\label{subsec:interpretation}
The Interpretation Stage is the main enhancement of our system over Cage's approach, integrating the \textit{I-Ching}'s semantic interpretations. 
From the coin casting, an original hexagram, any changing lines (Dong Yao), and the resulting transformed hexagram are formed.
We retrieve the corresponding context-free interpretations for each hexagram and line.
These data, together with the user's inquiry, are sent to \textit{Gemini 2.5 Flash} \cite{team2023gemini, comanici2025gemini}, a Large Language Model (LLM), via Google GenAI's API.
The inputs are prompt-engineered so that Gemini generates a reading tailored to the user's question, as well as keywords capturing mood, energy, dynamics, and spatial qualities.
The keywords are then passed to Lyria \cite{team2025live}, a text-to-music model in the Google GenAI suite.
Lyria generates 30–60 second ambient pieces reflecting the hexagram transformations in the context of the user inquiry.
A “Lyria Prompt JSON” button allows users to view the exact prompt used for music generation, providing transparency and reproducibility.

Alongside the generated music, the interface displays the hexagrams, changing lines, and textual interpretations as reference, grounding the auditory experience in the canonical \textit{I-Ching} structure.

\subsection{Sound Engine}
The system’s sound engine is implemented using the Web Audio API, with Tone.js providing abstractions for timing, routing, and synthesis. Sound generation operates in two stages: a real-time casting phase and a generative playback phase. During casting, each coin toss activates a loop-based layer corresponding to a hexagram line, rendered with sampler instruments derived from a soundfont library and routed through per-line panning into a shared compressor and reverb. The casting-stage music continues looping while the system generates the divinatory interpretation and music prompt, maintaining sonic continuity during computation. Once the interpretation is complete, the system transitions to a generative playback phase, where a language-model–derived music plan conditions a real-time text-to-music session using Google’s Lyria. At any point, a user-initiated reset can stop all active audio processes, disposes of looping layers, and returns the system to a silent initial state.

\subsection{User Interface}\label{subsec:userInterface}
The interface, as shown on the top of Figure~\ref{fig:flowchart}, is designed to prioritize minimalism and immersion, echoing the meditative and contemplative nature of the \textit{I-Ching} practice.
Users can backtrack, restart, or view instructions at any point through persistent navigation controls and tooltips. 
Each page features a breathing oracle circle in the background that slowly shifts in color and rhythmically expands and contracts.
The design draws inspiration from the character \textit{Samantha} in the film \textit{Her} and Apple's voice assistant \textit{Siri}, imitating an entity that listens and responds. 
To further reflect the cultural context of the \textit{I-Ching}, subtle overlays of Chinese ink painting textures are incorporated into the background.

Interactive elements are woven into the interface to engage users, including a chatbot-like inquiry system and user-initiated coin tossing. 
These interactions aim to make the experience personal and meaningful: the user's actions directly influence the result, emphasizing the concept that chance is co-created by the individual and the universe. 

\subsection{Process-driven Participatory Music Creation with AI}
Our system draws on a lineage in computer music that treats musical works as interactive processes rather than fixed scores. In their discussion of \textit{Computers and Future Music} \cite{maxmatthews}, Max Mathews et. al. describe a future in which listeners actively participate in shaping musical outcomes through interaction with computational systems. Our work can be understood as a contemporary realization of this model, implemented with generative AI algorithms.

Rather than influencing musical detail directly, participation in our system occurs through a ritualized question-and-casting interaction. Users pose a question and perform a coin casting according to the \textit{I-Ching}; the resulting hexagram initiates an interpretive process rather than prescribing musical structure. This aligns with Mathews’s conductor concept, exemplified by the GROOVE \cite{maxmatthews}, in which humans guide a computational process without specifying individual musical events.

In the interpretation stage of our system, a large language model generates a textual interpretation of the casting outcome and the user’s question. In this way, AI serves as an intermediate interpretive step: it generates a textual interpretation of the casting outcome, which is then used to condition the music generation process. The generative music model responds to this prompt to produce a musical realization.

The contribution of this work lies not in new synthesis or generation techniques, but in demonstrating how contemporary generative AI can be integrated into computer music systems as an interpretive interface rather than a compositional authority. By situating AI within a culturally grounded, participatory process, the system offers an alternative model for human–AI music interaction---one that emphasizes meaning-making and engagement over optimization or automation. This perspective extends early visions of participatory computer music into the context of modern AI-driven tools, and suggests new roles for language models in interactive music systems beyond direct control or replacement of human creativity.

\section{Discussions}
In developing this system, we encountered several intertwined aesthetic, technical, and cultural challenges. 
First, given the \textit{I-Ching}'s complexity---its many divinatory methods, centuries of philosophical commentary, and historical reinterpretations---no single work can capture it in its entirety. 
Because the \textit{I-Ching} continues to be widely used as a divinatory practice, we focus specifically on the popular approach of Wen Wang Fa as an interactive process.
The availability of large language models made it possible to generate interpretations that are specific to each user's consultation, and therefore supports a more unique, dialogic, and interactive experience.

Second, we spent considerable time brainstorming how to implement chance operations musically. Our goal was to preserve the unpredictability associated with Cage's practice while imposing constraints that maintain perceptible structure and continuity. The loop-based design, in which each coin row generates a probabilistic melodic layer, allowed us to keep the local musical rules simple yet produce a perceptible sense of accumulation. As more instruments enter, the sonic texture emerges in parallel with the hexagram, reinforcing the metaphor of lines gradually composing a pattern of meaning.

Finally, model controllability remains an open problem. Although Lyria is designed for conditioned, real-time music generation, we found that its responses were sometimes insensitive to nuanced changes in the textual prompts and did not always align with the intended mood or structural trajectory. This raises broader questions about the limits of current text-to-music systems in supporting fine-grained, semantically grounded control. Addressing these issues is crucial for deepening the connection between divinatory interpretation and sonic outcome.

Future work could extend this system toward community-driven forms that are more closely connected to users’ experiences. One direction is to aggregate anonymized user consultations and use recurring or divergent questions, along with their \textit{I-Ching} interpretations, as material for a generative artwork that sonifies collective patterns of inquiry. Another direction would introduce deeper user interaction by allowing participants to reflect on the moods elicited by an interpretation, with these affective responses informing adaptive music generation strategies. Together, these approaches would deepen the dialogic relationship between users, the \textit{I-Ching}, and the resulting music.

From a computer music perspective, this work shows how AI can be integrated into process-driven systems not as an autonomous composer or generator of finished artifacts, but as an interpretive intermediary within a participatory musical process. By foregrounding ritual, interpretation, and interaction, the system extends historical visions of participatory computer music into contemporary discussions on AI and meaning-driven musical systems.

\section{Conclusion}
The \textit{I-Ching} is foundational to Chinese philosophy and cosmology, and its divinatory practices continue to shape how people think about change, uncertainty, and decision-making. While Western experimental musicians such as John Cage have drawn inspiration from the text, their focus on chance has often bracketed out its semantic, ritual, and cultural dimensions. In \textit{Music of Changing Lines}, we attempted to re-balance that relationship by designing an immersive system that guides users through each stage of the Wen Wang Fa divination process and incorporates not only randomness but also meaning.

Our approach situates chance operations in the casting stage, where probabilistic musical processes mirror the unfolding of the hexagram, and then reintroduces semantic depth in the interpretation stage through LLM-driven textual readings and AI-generated music. In doing so, we move beyond treating the \textit{I-Ching} as a generic randomization tool and instead foreground it as a culturally situated, meaning-bearing framework.  
By embedding algorithmic and generative music within a participatory process, this work opens avenues for designing computer music systems that prioritize interaction, personalization, and the unfolding of musical experience over fixed artifacts.

\bibliography{main}

\end{document}